\documentstyle[prb,aps,epsf]{revtex}
\input{psfig}
\narrowtext
\begin{document}

\draft
\title{Local structure study about Co in
YBa$_2$(Cu$_{1-x}$Co$_x$)$_3$O$_{7-\delta}$
thin films using polarized XAFS}

\author{E. D. Bauer, F. Bridges,  C.  H.  Booth}
\address{
Physics Department,
University of California,
Santa Cruz, CA 95064 }

\author{J. B. Boyce}
\address{
Xerox Palo Alto Research Center, Palo Alto, CA 94304}

\author{T. Claeson, G. Brorsson}
\address{
Physics Department,
Chalmers Univ.  of Tech.,
S-41296 Gothenburg, Sweden}

\author{Y. Suzuki}
\address{AT\&T Bell Laboratories, Murray Hill, NJ 07974} 

\date{Draft: \today}
\maketitle

\begin{abstract}
We have studied the local structure around Co
in YBa$_2$(Cu$_{1-x}$Co$_x$)$_3$O$_{7-\delta}$ thin films 
with three different concentrations: x=0.07, 0.10, 0.17,
and in a PrBa$_2$(Cu$_{1-x}$Co$_x$)$_3$O$_{7-\delta}$ thin film
of concentration x=0.05 using the X-ray Absorption Fine
Structure (XAFS) technique.  Data were collected at the Co $K$-edge
with polarizations both parallel and perpendicular to the film
surface.  We find that the oxygen neighbors are well ordered
and  shortened in comparison with 
YBCO Cu-O values  to 1.80 \AA{} and 1.87 \AA{} in the $c$-axis 
and $ab$-plane, respectively. A comparison of further neighbors 
in the thin film and powder data show that  these peaks in the 
film are suppressed in amplitude relative to the powder samples, 
which suggests there is more disorder and/or distortions  of 
the Co environment present in the thin films. 
\end{abstract}



\section{Introduction}

\begin{figure}
\centerline{\hbox{\psfig{file=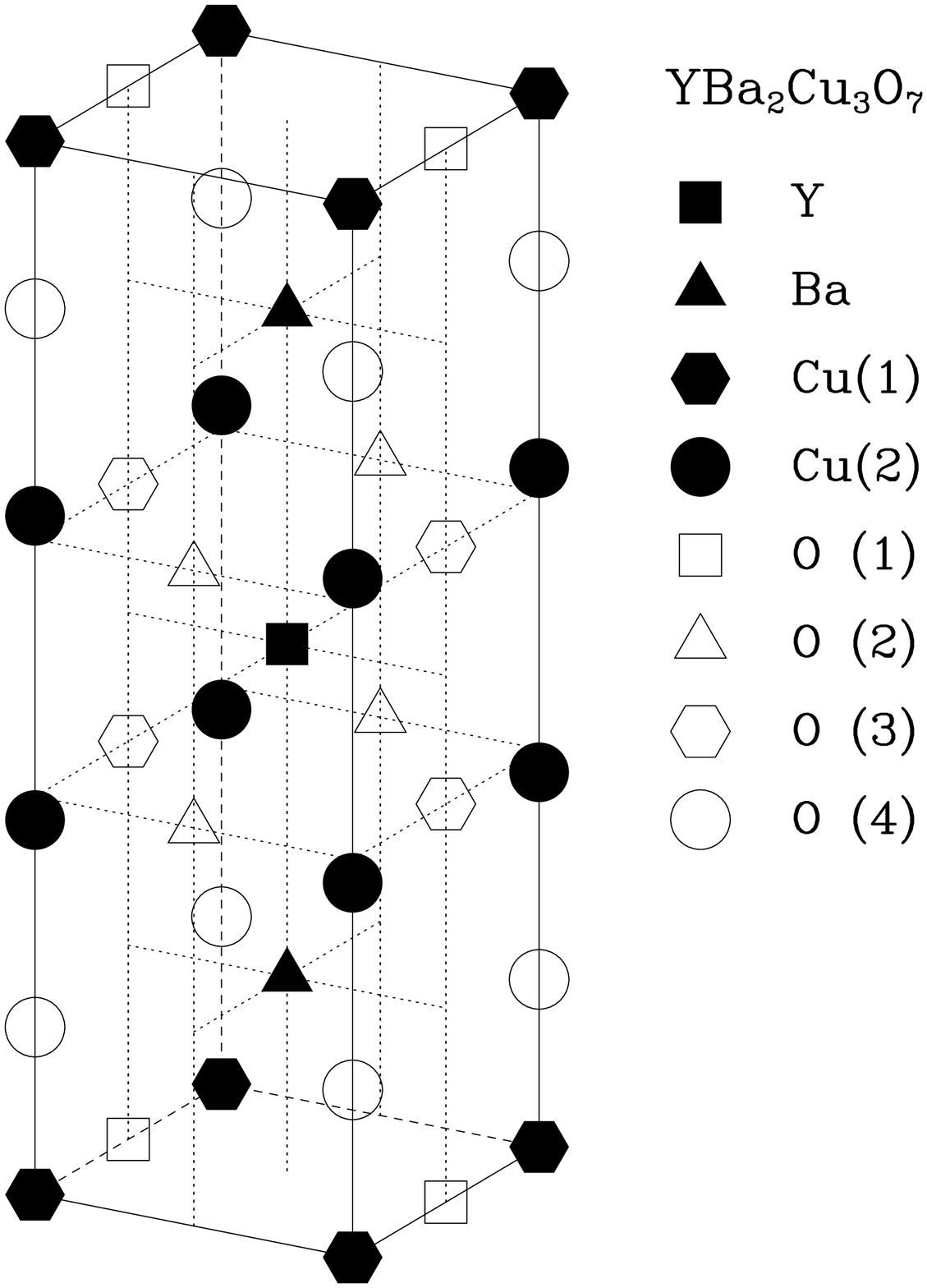,height=8in}}}
\caption{The crystal structure for YBa$_2$Cu$_3$O$_{7-\delta}$.  
Co is thought to substitute primarily on the Cu(1) site. The O(5) site
(not shown) lies halfway in between the Cu(1) sites along the $a$-axis.}
\label{YBCO_struc}
\end{figure}

The substitution of  different cations for copper
in YBa$_2$Cu$_3$O$_{7-\delta}$ (YBCO) has been widely used to
study the relationship between the crystal structure and superconductivity,
particularly the differences between Cu(2) in the CuO$_2$ planes 
and Cu(1) in the CuO ``chains'' (See Fig. \ref{YBCO_struc} for 
the crystal structure).  Previous investigations have shown that 
Co, Al, and Fe substitute primarily on the Cu(1) site at low defect
concentrations\cite{Bridges89,Kakihana90,Kakihana93}, 
and thereby provide a probe to study 
the effect of the Cu(1) chains on the superconducting properties. 
Co and Fe both suppress the superconducting transition temperature, $T_c$,
 to zero at roughly 15\% substitution in bulk 
samples\cite{Moeckly_unpublished,Zhao88}. 
However, the actual value of $T_c$ in a given sample depends  strongly on the
sample preparation. (For example we showed earlier that $T_c$
of  10\% YBCO:Fe thin film samples could vary 
from $\sim$ 0-80 K depending on the deposition temperature 
and cool-down rate\cite{Bridges92b}.) It has  also been 
observed\cite{Zolliker88,Yang90,Renevier93}
 that adding Co (or Fe) to YBCO increases the O concentration. 

Many experimental reports indicate that the local environment about Co is
strongly distorted or disordered. M\"{o}ssbauer experiments suggest that 
there are 3-4 inequivalent sites for Fe (Co  should be similar)
with different O coordinations\cite{Blue88}.
X-ray Absorption Fine Structure (XAFS) experiments 
show that the first neighbor oxygens are quite well ordered but the 
further neighbor  environment is distorted\cite{Bridges89,Renevier93,Li93}. 
Several groups have 
interpreted their data in terms of chain-like Co clusters in the Cu(1) 
layer\cite{Moeckly_unpublished,Renevier94b}
and in some cases a $<$110$>$ off-center displacement of some of the Co
atoms\cite{Bridges89,Li93,Renevier94a}. These small clusters are thought 
to be responsible 
for the suppression of  $T_c$ by distorting the Cu planes or by removing 
holes from the superconducting layer.
Each of these reports has a different proposal for the local structure,
again suggesting that it is heavily dependent on sample preparation. 
There is also the question\cite{Martin95} as to whether all the Co 
is actually in the bulk YBCO  -- is some Co 
located in grain boundaries or within other small crystallites?

Recent interest in Co substituted YBCO (YBCO:Co) has arisen in conjunction 
with developments in  SNS (Superconducting-Normal-Superconducting)
junctions. In these junctions it is highly desirable for the normal metal
to have the same thermal expansion and lattice parameters as YBCO. YBCO:Co
is nearly ideal in this respect and produces junctions with the lowest
resistance\cite{Char94,Antognazza95,Koren94}. 
Despite the enthusiasm for these devices and the finding that
YBCO:Co makes a good normal conductor for the junctions, there have been 
only been a few investigations involving YBCO:Co thin 
films\cite{Char94,Antognazza95,Koren94,Song91}. 
Previous studies of Co substituted YBCO have been on powder samples for which
the relative contributions of the various peaks in the XAFS $r$-space plots are
different than in polarized XAFS studies. 

In this paper, we present a polarized XAFS study of the 
local structure around the Co ions in thin films of
YBa$_2$(Cu$_{1-x}$Co$_x$)$_3$O$_{7-\delta}$.
 Three  Co concentrations
were used: x=0.07, 0.10, 0.17, and the $T_c$'s were similar to bulk material.
We find that the
nearest neighbor oxygen atoms about the Co are contracted both along the
$c$-axis and in the $ab$-plane.  In previous XAFS measurements\cite{Bridges89} 
on powder
samples, an average contraction was thought to be primarily a result of
an $ab$-plane contraction.  We also studied a 
PrBa$_2$(Cu$_{1-x}$Co$_x$)$_3$O$_{7-\delta}$ (PBCO:Co) thin
film sample with a Co concentration of x=0.05 and $T_c$ = 0, to see
if any significant change occurred in the Co environment when Y is 
replaced by Pr.

A brief introduction to polarized XAFS is presented in 
Sec. \ref{xafs}. In Sec.\ref{experiment} we 
give some experimental details and then 
outline the general features of the data, including a 
comparison with powder data in Sec \ref{results}.  A
discussion of theoretical simulations for various Co distortions
 is contained in  Sec. \ref{simulation}. 
We describe our fits in Sec. \ref {detailed-fits}
and discuss our conclusions in Sec \ref{conclusion}.

\section{Polarized XAFS}
\label{xafs}

The standard XAFS equation for a polycrystalline sample can be written as:
\begin{equation}
k\chi(k)=\sum_{j}\frac{A_{j}(k)}{R_{j}^2}
sin[2kR_{j}+\phi_{j}(k) ] exp(-2k^{ 2}\sigma^{2}_{j}- 2R_j/\lambda(k))
\end{equation}
where the
sum is taken over the shells of atoms at a distance $R_{j}$ from the absorbing
atom, with amplitude $A_{j}$ ,$k$ is the  $k$-vector of the ejected photon 
($k=[2m(E-E_{o})]^{1/2}/\hbar$ where E is the photon energy, $E_o$ is the edge
energy), $\phi_{j}(k)$ is the total phase shift of the photoelectron 
due to its interaction with the back scattering and absorbing atoms,
$\lambda$ is the electron mean free path, 
and $\sigma_{j}$ (Debye-Waller factor)  is the mean variation of $R_j$
arising from static and thermal disorder.  The amplitude is given
by:
\[
A_{j}= N_j S_0^2 F(R_j,k) 
\]
in which $N_j$ is the number of like atoms at a distance $R_j$, $S_0^2$
is an overall reduction factor accounting for shake-up and shake-off
processes, and $F(R_j,k)$ is the scattering amplitude.
The XAFS technique and data analysis process have been described in detail
elsewhere (for example, see Ref. \onlinecite{Hayes82,Teo86}).

In the absorption process the outgoing electron is preferentially ejected along
the
x-ray polarization vector, $\hat{\bf P}$, and the outgoing
electron intensity distribution varies as
$cos^2(\theta)$ = ${({\bf \hat R}{\bf \cdot\hat{ P}})^2}$ where
$\bf{\hat R}$ is in the direction of
the ejected electron. Consequently, for non-cubic materials one can take
advantage of the high degree of polarization of synchrotron radiation by
using an
oriented sample, usually a single crystal or oriented thin film. Neighboring
atoms along the polarization direction are preferentially probed while atoms
located in a plane perpendicular to $\bf{\hat P}$ are not observed.
Using polarized XAFS in the context of YBCO:Co
 has the advantage of mimizing the contributions from the
in-plane O(1) atoms with the x-ray polarization along the
$c$-axis of the film (i.e. normal to the film); conversely,
when the polarization vector
is in the $ab$-plane, the contribution from the axial oxygen O(4)
atoms is negligible.

\section{Experimental details}
\label{experiment}

The YBCO:Co films were made by pulsed laser deposition onto SrTiO$_3$
substrates and are estimated
to be 3000 \AA{} thick. Transition temperatures for the three concentrations
of Co in YBa$_2$(Cu$_{1-x}$Co$_x$)$_3$O$_{7-\delta}$ 
are: 38 K for x=0.07, 42 K for x=0.10, and
nonsuperconducting for x=0.17.  The PBCO:Co sample was 
prepared by laser ablation on a  LaAlO$_3$ substrate and is 5000 \AA{} thick.
All of the films are aligned with the $c$-axis perpendicular to the surface. 

X-ray absorption spectra were
obtained at the Stanford Synchrotron Radiation Laboratory 
(beamline 4-3) using a Si(220) monochromator crystal.  For each sample,
several traces 	were collected at 80 K in fluorescence mode using a 13-element 
Ge detector. In each case, data were collected with the x-ray polarization
vector approximately parallel or perpendicular to the film surface (within 
9 degrees; the error in cos$^2$($\theta$) is less than 3\%) 
in order to probe 
different atoms within the structure. 
Count rates were kept below 5 x 10$^4$/sec and corrected for the dead-time 
($\sim 5-10 \mu s$) caused by the energy-resolving spectroscopy amplifier.  
An important use of the 
Ge detector was to remove Bragg spikes occurring at the incident beam energy
in some of the channels.  These spikes were removed by replacing them
with the normalized  average  data for the other detector elements.  

\section{Experimental results}
\label{results}
\subsection{Qualitative features}
\label{Qualitative}
\begin{figure}
\centerline{\hbox{{\psfig{file=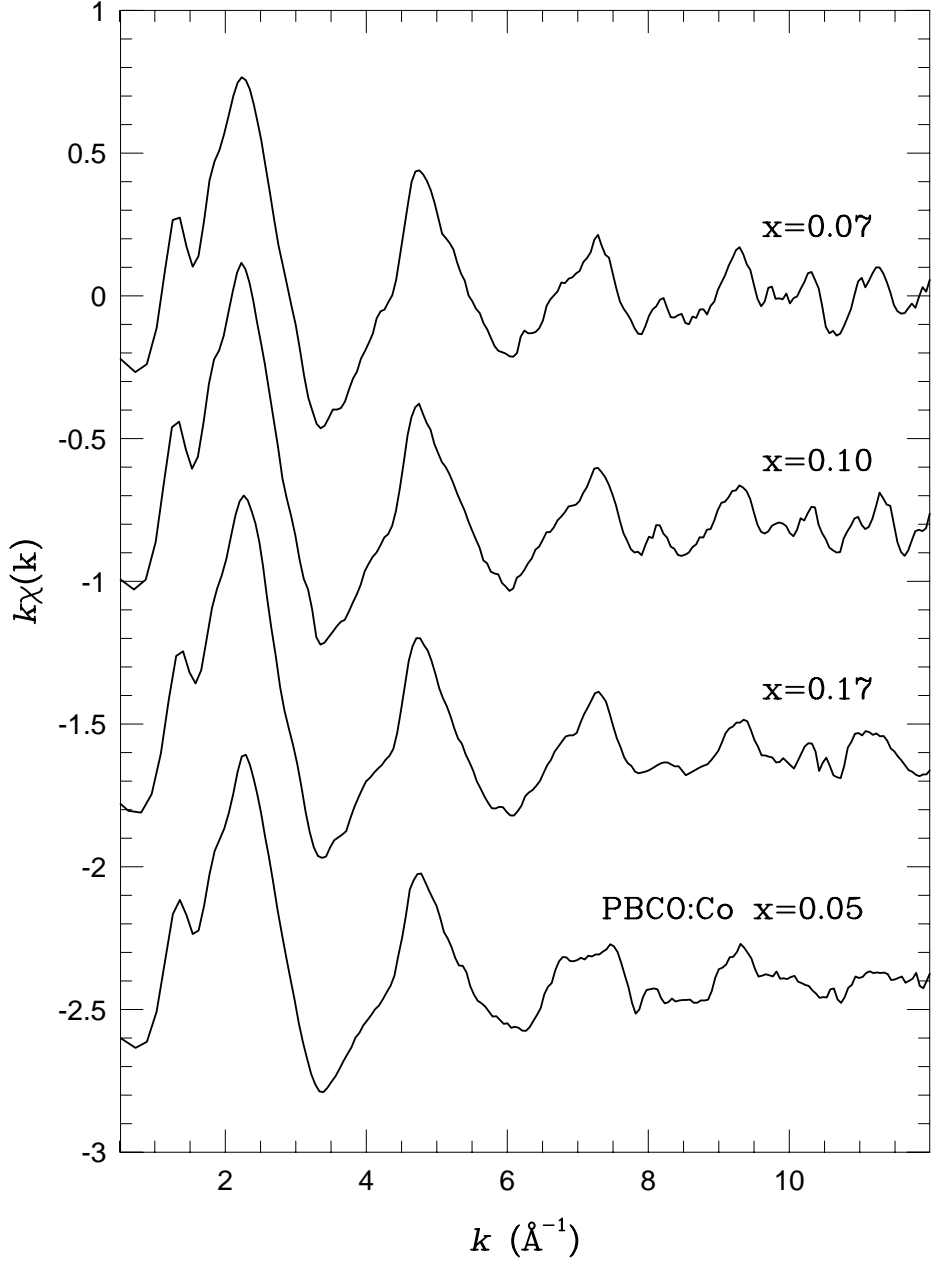,width=6in}}}}
\caption{$k \chi(k)$ vs. $k$ for the Co $K$-edge
 with the polarization along the $c$-axis
 for the three concentrations of 
YBa$_2$(Cu$_{1-x}$Co$_x$)$_3$O$_{7-\delta}$ and the
PrBa$_2$(Cu$_{1-x}$Co$_x$)$_3$O$_{7-\delta}$ 
sample.  The three curves below the top one have been displaced
for clarity.}
\label{kspace_c}
\end{figure}

\begin{figure}[th]
\centerline{\psfig{file=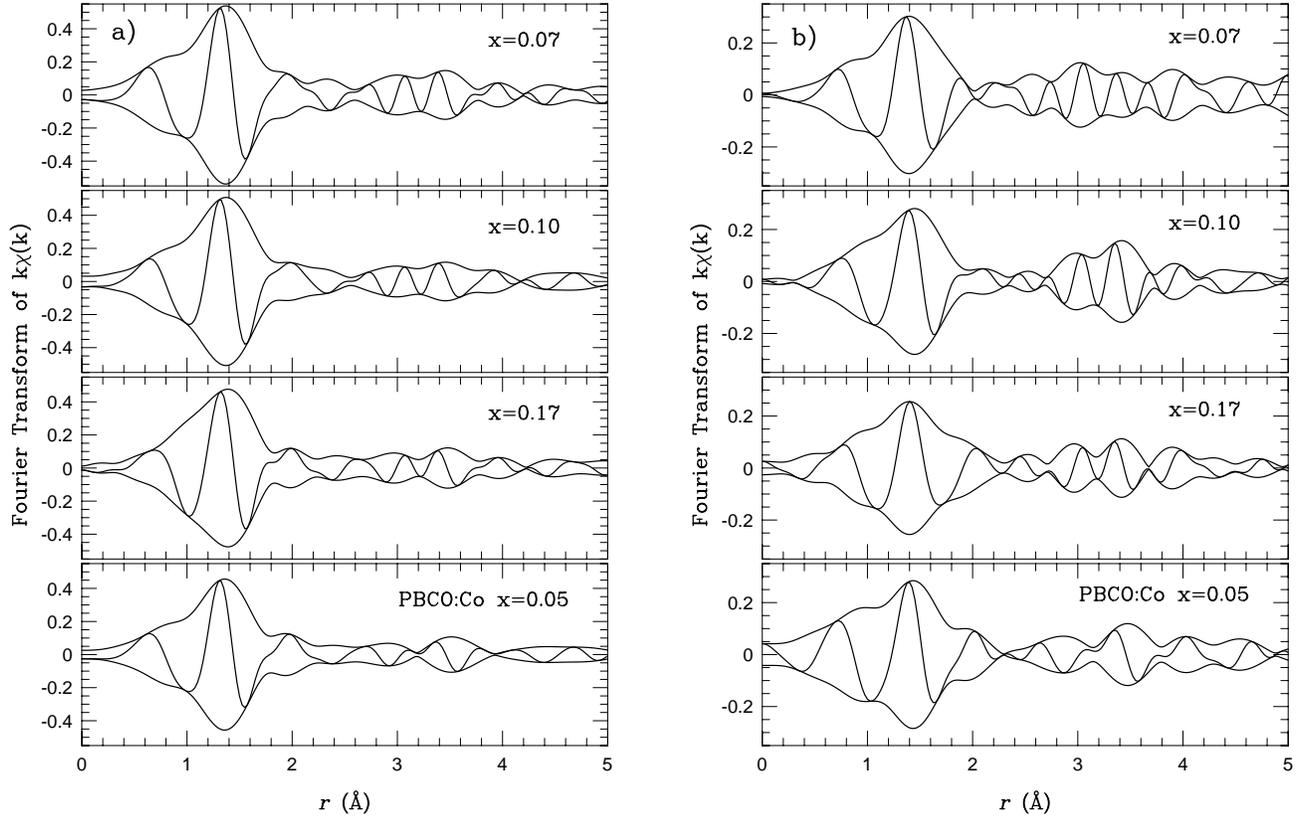,width=7in,rheight=8in}}
\vspace{-3in}
\caption{ The left panel a) is the Fourier Transform of 
the $c$-axis Co $K$-edge data for the three
YBCO:Co concentrations with the PBCO:Co data on the bottom.  
The right panel b) shows the Fourier Transform of the $ab$-plane data
for the same samples. The envelope curves are the  magnitude of the
transforms and the fast oscillatory curves are the real parts of the
transforms.  For each data set the transform
range is from 3.5 to 11.5 \AA$^{-1}$ and Gaussian broadened by
0.3 \AA$^{-1}$.}
\label{polar_data}
\end{figure}

An example of the $k$-space data, $k\chi (k)$, for the four samples studied 
is shown in Fig. \ref{kspace_c}, with the polarization vector parallel 
to the $c$-axis. 
The Fourier Transform (FT) of these XAFS data 
is shown in Fig. \ref{polar_data}a. Peaks in the $r$-space data correspond 
to neighbors at different distances with a phase shift, typically -0.4 to 
-0.2 \AA{}, due to the photoelectron interaction with the potentials 
of both the absorbing and the neighboring atom. 
For these data, the first peak at $\sim$ 1.4 \AA{} corresponds to 
two O(4) neighbors along the $c$-axis, and is well ordered. 
 However, the further neighbor peaks are very small in amplitude,
much smaller than the corresponding peaks for Cu XAFS of an undoped
 high quality YBCO film.  There should be large
contributions from the Ba and Cu(2)/(Co(2)) neighbors in the 3-4 \AA{} range;
the absence of these well-defined peaks in the $c$-axis polarization data
suggests a large amount of distortion and/or disorder in the Co further 
neighbors.  This distortion/disorder 
appears to be similar for the three YBCO:Co samples but increases with 
Co concentration. In addition, the PBCO:Co sample also  has a large 
distortion of the further neighbors even at only 5\% Co.

The corresponding $r$-space plots for the $ab$-plane data are shown in 
Fig. \ref{polar_data}b. Again, the first O peak (Co-O(1)/O(5)) is 
well defined (in 
this case it is an average within the $ab$-plane since the film is twinned 
and the amplitude is therefore reduced by a factor of two.) 
For this polarization,
the further neighbor contributions in the 3-4 \AA{} range should arise 
from the Ba and Cu(1)/Co(1) atoms.

\begin{figure}[th]
\centerline{\hbox{{\psfig{file=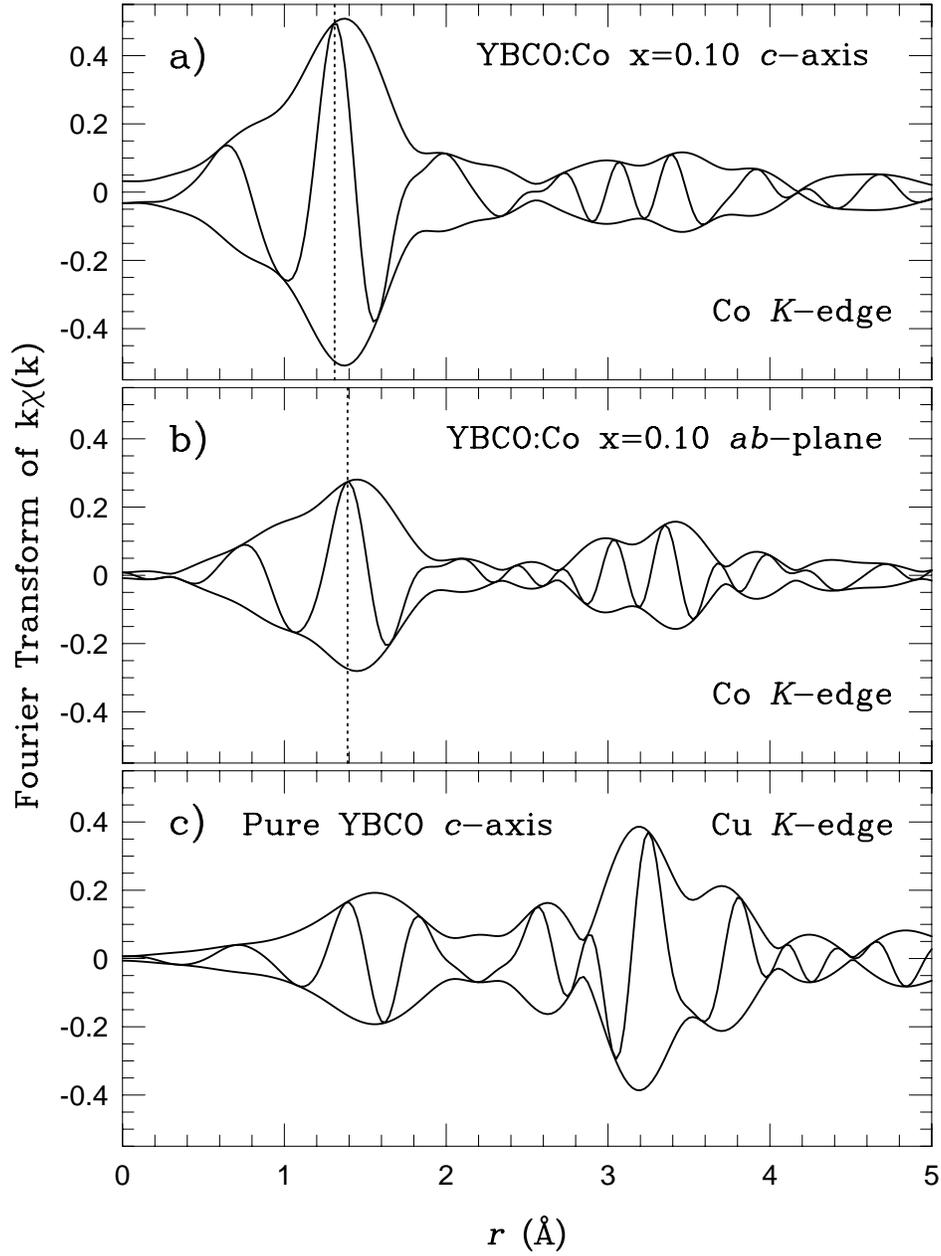,width=6in}}}}
\caption{A comparison of the YBCO:Co with x=0.10 of the a) $c$-axis
Co $K$-edge data and the b) $ab$-plane Co $K$-edge data.  
$c$-axis polarized Cu $K$-edge data from normal YBCO
is shown in c).  For each data set the transform
range is from 3.5 to 11.5 \AA$^{-1}$ and Gaussian broadened by
0.3 \AA$^{-1}$.}
\label{comparison_co10}
\end{figure}

A comparison of the data (Fig. \ref{comparison_co10}) for the two 
orientations shows immediately (dotted lines in Fig. \ref{comparison_co10})
 that
the $c$-axis and $ab$ plane O atoms are at different distances by roughly 
0.1 \AA.  This means that the Co is not in a cubic environment and  
the structure containing the Co is epitaxial with the substrate 
and  the YBCO; hence, if Co were in a defect phase it would therefore 
have to be non-cubic and oriented with the substrate, which is unlikely.
It is consistent with the Co replacing Cu
in YBCO but with the CoO$_z$ cluster displaced from the usual Cu(1) site. 
The $c$-axis O peak is primarily a Co(1)-O(4) peak (a Co(2)-O(4) bond 
length is expected to be much longer) and the fact that it is shorter 
than the Co-O(1) in the $ab$-plane supports the previous conclusion 
that Co substitutes primarily on the Cu(1) site.
A comparison with $c$-axis polarized Cu $K$-edge pure YBCO film data 
(Fig. \ref{comparison_co10}c; $ab$-plane data not shown)
 indicates that the Co-O peaks are shorter in both directions than the 
corresponding Cu-O peaks in the YBCO thin film; therefore, the O cluster 
about Co is contracted compared to that about Cu.  In comparing the $c$-axis
data note that for Co(1), there are two O(4) neighbors; for Cu there are
$2/3$ ($1/3$ of Cu is from Cu(1)$\times$ 2 O(4) neighbors) at a 
distance of $\sim 1.4$ \AA{} and $2/3$ neighbors ($2/3$ of Cu is from 
Cu(1)$\times$ 1 O(4) neighbor) at a longer distance of $\sim 1.9$ \AA{}.

\subsection{Comparison with data for powder samples}
\label{powder}

\begin{figure}[th]
\hbox{\centerline{\psfig{file=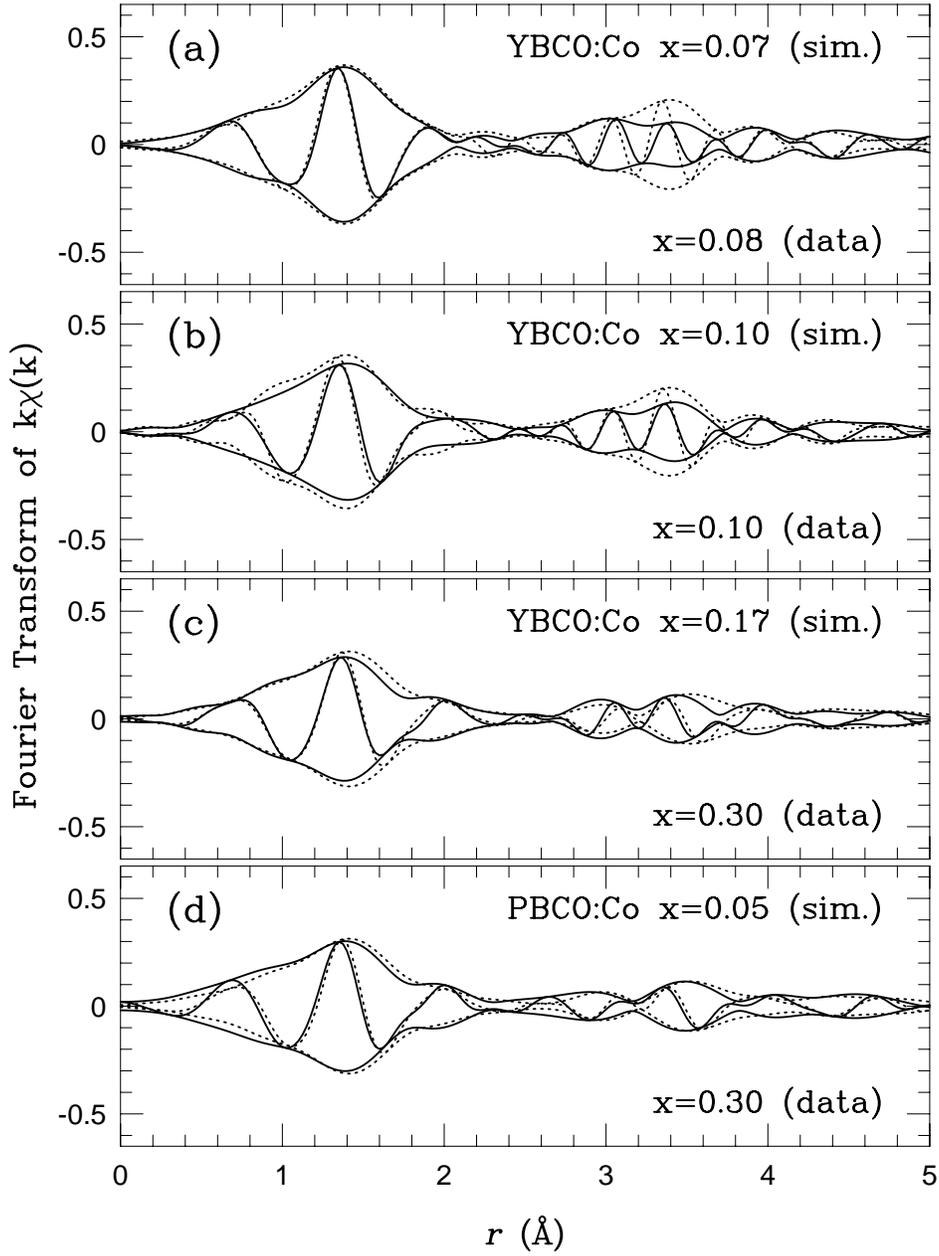,width=6in}}}
\caption{ a) Comparison of YBCO:Co with x=0.07 
$c$-axis and $ab$-plane Co $K$-edge data added
together in the proportions 2/3($ab$-plane) + 1/3($c$-axis)
 (solid) to that of an 8\%
powder sample (dotted); b) summed 10\% film data (solid)
 and 10\% powder data (dotted); c) summed 17\% film data (solid)
 and 30\% powder data (dotted); d) the
summed PBCO:Co film data (solid) is most similar to 
the 30\% powder data (dotted).
For all panels, the $k$-window is 3.5 to 11.5 \AA$^{-1}$ and Gaussian
broadened by 0.3 \AA$^{-1}$. }
\label{comp_pwd}
\end{figure}

To compare
the present results with earlier YBCO:Co powder XAFS data\cite{Bridges89}, 
the $c$-axis 
and $ab$-plane data for the thin films have been added together in the correct  
proportions (2/3($ab$-plane) + 1/3($c$-axis)) to simulate a random 
orientation -- i.e. a powder.  (See Fig. \ref{comp_pwd} for a comparison 
of the resulting FT spectra.)  There are similar Co concentrations for 
some of the YBCO:Co powder samples: for x=0.07 (x = 0.08 for powder) and for 
x=0.10 (same for powder). However there is no corresponding x = 0.17
powder sample. 
For this concentration, the film data seems most similar to the
x=0.30 powder sample.
The x=0.05 PBCO:Co data is more disordered in the further neighbors than 
YBCO:Co and also compares well with the  x=0.30  powder data.

A comparison of the simulated  powder data with our earlier powder data shows
that the nearest neighbor oxygen environment around the cobalt is very similar 
in each case. For the further neighbors, the positions of the 
oscillations of the real part of
the transform match well; however, the amplitude of the further neighbor 
peaks for the thin film  and powder samples differ. 
Thus,  the Co atoms appear to have similar distortions in powder and 
thin film samples, but the films show much more disorder.
It is interesting to note that the 17\% (film) data has the same amount of 
disorder as in a powder sample of nearly twice its concentration 
(in which most of the Cu(1) atoms 
have been  replaced by Co) and supports the suggestion that Co forms clusters 
in the Cu(1) layer. This effect is even more pronounced
in the 5\% PBCO:Co film which has a comparable degree of disorder to that
of the 30\% powder sample.  A large amount of disorder caused by
such small concentrations of Co in the thin films 
may indicate larger Co clusters 
have formed in the 17\% film which is similar to the structure when most
of the Cu(1) is replaced by Co in the powder samples. 
The nature of the distortions and the
amount of disorder are most likely sample dependent; however both the YBCO:Co 
and PBCO:Co films, which were grown in different laboratories, 
have more disorder at low Co concentrations than the corresponding bulk 
material.  This might be 
evidence that the Co environment in thin film samples is inherently more 
disordered than in the bulk materials.

\section{Simulated models for possible Cobalt distortions}
\label{simulation}

The disorder and/or distortions in the further neighbors 
(Fig. \ref{polar_data}) have  so far proved too complicated to fit reliably.
Therefore, we have calculated a number of theoretical XAFS simulations 
at a temperature
of 80 K using the FEFF6 code\cite{FEFF5,FEFF6}
in order to determine some possible distortions which  might lead to the
reduced further neighbor peaks. 
In these calculations, we first use the known crystal structure of YBCO to 
provide an approximately correct local geometry, and make separate 
calculations for Co on the Cu(1) and Cu(2) sites. This generates a 
simulation of the FT spectrum and provides a starting point for 
refinements of the local structure discussed below. If no
distortions of the Co(1) or Co(2) sites occurred, the further neighbor peak
amplitudes would be comparable to that observed for polarized Cu $K$-edge
studies of pure YBCO thin films.  As noted above, this is inconsistent with 
the data (see Fig. \ref{comparison_co10}c). We next consider
several possible distortions for Co on the Cu(1) site. First, we use the
distortions obtained in earlier studies\cite{Bridges89,Li93} of bulk 
powder samples which suggested
that a fraction of the Co are displaced off-center along the $<$110$>$
direction by $\sim$ 0.2 \AA; the fraction depended on the type of Co chain-like
clusters that formed. We also include a small contraction of both nearest
neighbor Co-O bond distances. Using typical Debye-Waller parameters,
the Co further neighbor composite-peak for this distorted site are
significantly larger than observed in the film data. For the $c$-axis data, the 
simulations show that the Co(1)-Ba peak is considerably reduced, but a 
large Co(1)-Cu(2) peak (which includes multi-scattering from 
Co(1)-O(4)-Cu(2)) occurs near 3.7 \AA{} in the $r$-space data, that is 
not significantly decreased by this $ab$-plane 
distortion. Consequently, to explain the lack of such a peak in the $c$-axis
data, requires that some other distortion/disorder be present. 
The O(4) atoms may be displaced such that the Co(1)-O(4)-Cu(2) unit is
not co-linear, thus reducing the multi-scattering contribution\cite{Teo86},
 or the Co might
be displaced along the $c$-axis. We could alternatively use a large 
Debye-Waller parameter, $\sigma$; to suppress the Co-Cu(2) peak 
sufficiently would require $\sigma$ $\sim$ 0.1 \AA. 

\begin{figure}[th]
\centerline{\hbox{\psfig{file=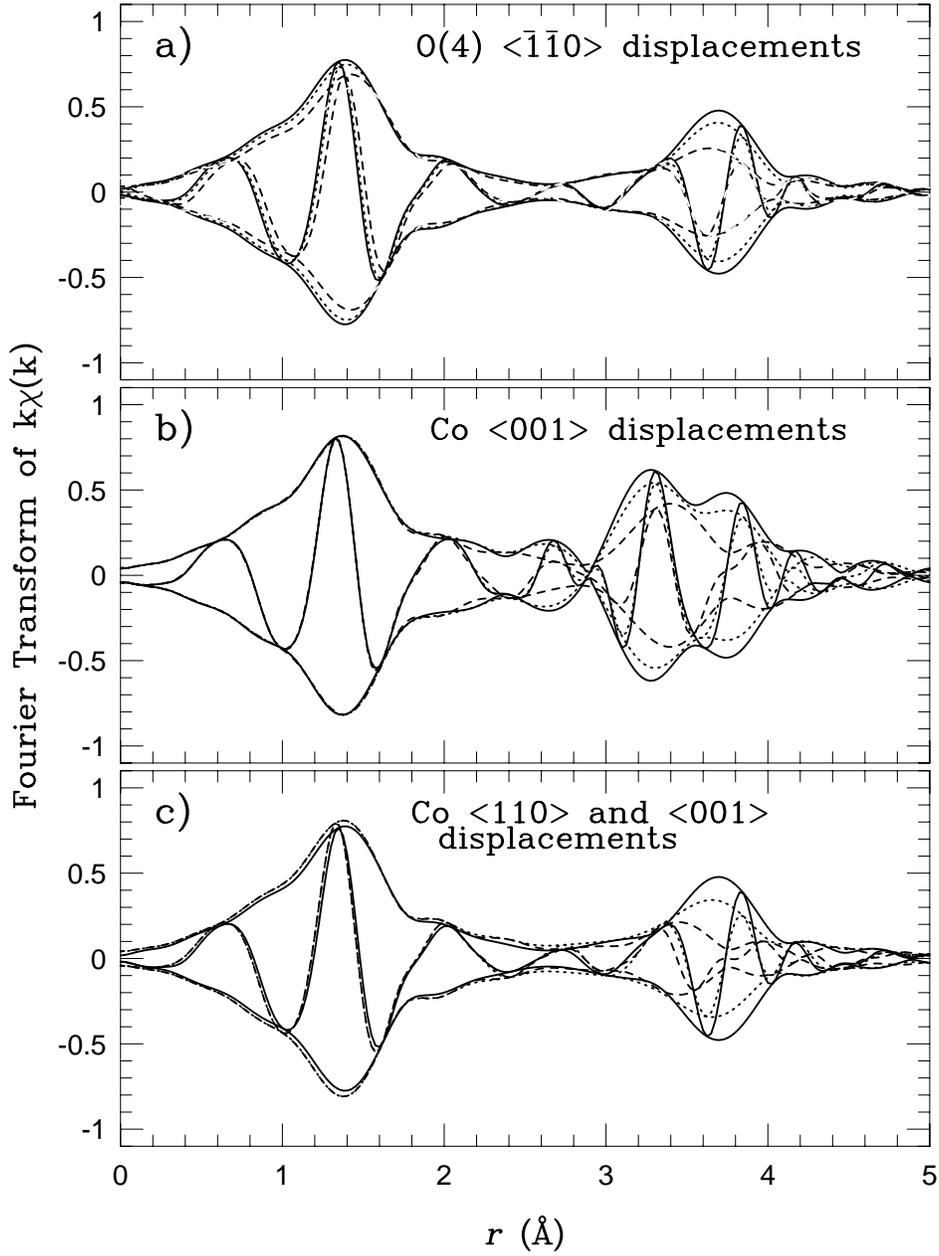,width=6in}}}
\caption{ FEFF simulations at $T$=80 K for the Co $K$-edge 
with polarization vector along the $c$-axis.
a) illustrates various O(4) displacements: 
O(4) not moved from its original site (solid),
O(4) displaced by  0.1 \AA{} (dotted), and O(4) displaced by 
 0.2 \AA{} (dashed) in the $<\overline{1}\overline{1}0>$ direction.
The Co(1) atom is distorted by
 0.2 \AA{} in the $<$110$>$ direction in each case.
b) shows simulations of various displacements of the Co(1) 
along the $<$001$>$ direction:  Co(1) in the ideal Cu(1) site (solid),
and Co(1) displaced by  0.05 \AA{} (dotted) and  0.10 
\AA{} (dashed).  c) shows simulations in which the Co(1) atom is
displaced  0.2 \AA{} in the $<$110$>$ direction (solid) as in a) 
 but with an additional displacement along the $c$-axis:  
Co(1) displaced by  0.05 \AA{} (dotted) and  0.10 
\AA{} (dashed). In the last two panels, the Co-O(4) distances
 have been shortened
by  0.05 \AA{} for a better simulation of the film data. 
For each data set the transform range is from 3.5 to 11.5 \AA$^{-1}$.}
\label{simulation_fig}
\end{figure}

For the first type of distortion, with the Co displaced
$\sim$ 0.2 \AA{} in the $<$110$>$ direction, perfect co-linearity would
require the O(4) to be displaced $\sim$ 0.1 \AA{} in the $<$110$>$ direction.
However, the difference in the simulation between perfect co-linearity and 
an undisplaced O(4) is small. We then consider a number of O(4) 
displacements (eg. 0, 0.1, 0.2 \AA{} ) in the  
$<\overline{1}\overline{1}0>$ direction to further decrease the 
multi-scattering contribution. 
The O(4) atoms must be displaced  in this way
by $\sim$ 0.2 \AA{} (Fig. \ref{simulation_fig}a) from its original
site in order to have an amplitude of the Co(1)--Cu(2) peak
comparable to what is found experimentally. It is not clear what 
would drive such a distortion in this crystal. 

If the Co were distorted only along the $c$-axis, the Co-Ba would still be 
visible, although slightly reduced in amplitude, as shown in 
Fig. \ref{simulation_fig}b (the Co-Ba contribution has
a peak near 3.3 \AA). Such a distortion would produce two Co-Cu(2) peaks which
could destructively interfere.  To reduce the Co(1)-Cu(2) peak 
amplitude (at 3.7 \AA) in this way to be comparable with the thin film 
data would require a displacement of the Co by
$\sim$ 0.1 \AA. In order to reduce both the Ba and Cu(2) peaks, a combination 
of the above Co displacements (both in the $<$110$>$ and 
$<$001$>$ directions) may be necessary. In  Fig. \ref{simulation_fig}c  
we show simulations with the Co displaced $\sim$ 0.2 \AA{} in the 
$<$110$>$ direction as before, but with a variety 
of displacements (eg. 0, 0.05, and 0.10 \AA{}) in the $<$001$>$ direction	
  as well.

The argument presented previously for a $<$110$>$ Co distortion is based on the
contracted Co-O bond length in the $ab$-plane. If the two nearest O neighbors 
are on the $a$ and $b$ axes, the short Co-O bonds  support the conclusion
that the Co atoms have a $<$110$>$ off-center displacement. 
The possibility of a $c$-axis displacement is less clear. 
Since the CoO$_z$ cluster is contracted, it may be that
the cluster is unstable on the Cu(1) site and moves towards one of the
neighboring Cu(2) atoms.

Finally, we note that simulations for the $ab$-plane data show that a $<$110$>$ 
Co distortion reduces both the Co(1)-Ba and the Co-Cu(1)/Co(1) contributions, 
but for a $\sim$ 0.2 \AA{} distortion, the suppression is not as large as 
observed in the film data. $c$-axis displacements will reduce the Co(1)-Ba
neighbor peak further but not the Co-Cu(1)/Co(1) peak.

\section{Detailed fits and discussion}
\label{detailed-fits}

For detailed fits of the data we used theoretical standard functions for the 
various atom-pairs (Co-O, Co-Ba etc.) calculated using the
FEFF6 code. In our previous study\cite{Bridges89,Li93}
 of YBCO:Co bulk samples, we proposed that the cobalt
ions formed short zigzag chains along the $<110>$ direction in the Cu(1)
layer.  $<110>$ displacements of some of the Co could produce a long Co-O peak
(at 2.4 \AA{}), 3 Ba peaks, one at the normal distance
and two separated by $+/- \delta r$ from it, and a shortened
Co--Cu(1)/Co.  Full fits to the nearest neighbors out to 4 \AA{}
 to both the $ab$-plane and $c$-axis 
orientations using this distortion were tried as well as to
other distortion models (such as Co displaced a fixed amount in 
the $<110>$ and then distorted
along the $<001>$ direction,  or Co displaced in the $<100>$ direction, etc.)
 all with similar results:  fits to the further neighbors were of poor
quality.  Thus, we have limited our fits to only the first neighbor
O atoms.

\begin{figure}[th]
\centerline{\hbox{\psfig{file=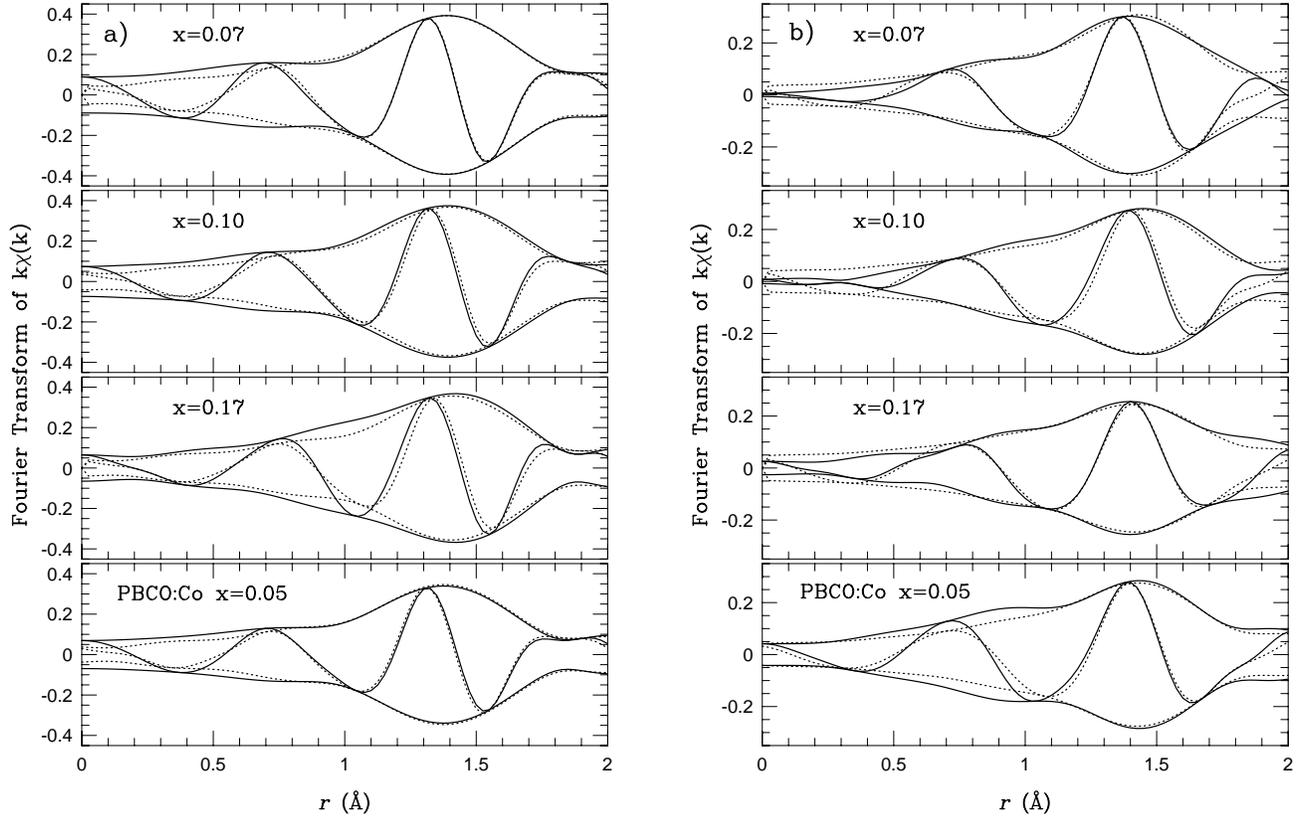,width=7in,rheight=8in}}}
\vspace{-3in}
\caption{ a) single peak fits (dotted) to the Co-O(4) atom-pair
for the $c$-axis data (solid). The fit range is from 1.2 to 1.8 \AA{}.
The $k$-window for the fits
is 4.5 to 11.5 \AA$^{-1}$ and Gaussian
broadened by 0.3 \AA$^{-1}$. 
b) single peak fits (dotted) to the Co-O(1) atom-pair for the $ab$-plane 
data (solid).  The fit range is from 1.2 to 1.8 \AA{}.  The $k$-window for 
the fits is 3.5 to 11.5 \AA$^{-1}$ and Gaussian
broadened by 0.3 \AA$^{-1}$. }
\label{fits}
\end{figure}

For the $c$-axis data, we used a FT range of 4.5-11.5 \AA$^{-1}$ 
and only fit the first neighbor O peak, either to a single peak or to a sum 
of two peaks. Generally the single peak fit ($r$-space fit range  1.2-1.8 \AA) 
was good, but there were a couple discrepancies -- the Debye 
Waller parameter was 
too small in some cases and the deviation between the fit and the data
 was worse near the higher part of the fit range. 
The single peak fit results are summarized in
Table I and the fits to the Co-O(4) peak are shown in Fig. \ref{fits}a.
The O(4) distribution is consistently short at a distance of $\sim$ 1.80 \AA{} 
(YBCO has a Cu(1)-O(4) bond length of 1.86 \AA). 
For the two peak fit we extended the $r$-space fit range
slightly to 1.2-2.0 \AA, and obtained a significantly better fit 
(this fit matched the main part of the $r$-space peak better); however,
the second peak which occurs at a longer distance of $\sim$ 2.3 \AA{}
 has a small amplitude of perhaps  20\%.  This longer O(4) peak
might be consistent with a Co(2)-O(4) bond.  However, 
since the second peak in the fits overlaps the tail of the Co-Ba peaks it is 
not clear that this O peak is significant. 

For the $ab$-plane data, the highest quality fits to the 
Co-O(1) distribution (Fig. \ref{fits}b) were obtained using a $k$-window of
3.5-11.5 \AA$^{-1}$ and a fit range of 1.2 to 1.8 \AA. 
This peak is well ordered,
although for the x=0.07 sample a  one peak distribution does not
fit well.  Another peak at $\sim$ 2.3 \AA{} improves the quality of the fit
but without a reasonable fit to the further neighbors, it is difficult to 
conclude at the present time whether this peak
is real or if it is only correcting in a crude way for the tail of the 
further neighbor peaks.

Our results of short, well-defined Co-O peaks, yet disordered further
neighbors, for both the $c$-axis and
$ab$-plane YBCO:Co and PBCO:Co data indicate the prescence of CoO$_z$ clusters.
The average distance of this first neighbor shell is 
1.83 \AA{} \((1/2\times1.79 + 1/2\times1.86=1.83\) \AA{}) and increases slightly with
increasing Co concentration.  If it is assumed that Co has an overall
reduction factor, S$_o^2$, of 0.7, then there are $\sim$ 2 O(4) and
$\sim$ 2-3 O(1) neighbors. These results are very similar to those
found in previous measurements on YBCO:Co 
powders\cite{Bridges89,Renevier93,Li93}. 
The reduced amplitude of the
Co-Ba and Co-Cu(1)/Cu(2) peaks observed in our data is indicative of
Co(1) being displaced from the (000) position rather than just  
some other distortion in these atom pairs, although both situations are
possible.  Off-center distortions of the Co atoms are
consistent with our earlier work\cite{Bridges89,Li93}
 as well as that of
 Renevier {\it et al.}\cite{Renevier93,Renevier94d}, 
but in disagreement with
others who found no distortions\cite{Yang90,Padalia92}.

\section{Conclusion} 
\label{conclusion} 

We have investigated the local structure about Co in YBCO:Co and
PBCO:Co  thin films 
and compared our results with measurements on bulk powder samples. The 
nearest neighbor oxygen atoms are
well ordered in both data sets which is evidence of the formation
of tightly bound CoO$_z$ clusters. The Co-O distances along the
$c$-axis and within the $ab$-plane are distinctly different and shortened 
compared to similar Cu-O bonds in YBCO in both 
cases: 1.80 and 1.88 \AA{} respectively. Thus, the Co substitutes
in a region that is epitaxial with the $c$-axis oriented YBCO. 
The short Co-O bond 
distance for the $c$-axis polarization supports the Co(1) substitution
site, but a small fraction, up to 10-20\%, may be on the Cu(2) site. 
The further 
neighbors peaks  are decreased in amplitude in the films relative to 
powder samples, 
indicating that the Co environment in these films is  more 
disordered or distorted than in 
the powder samples.  Theoretical simulations suggest that the Co
could be distorted along the $<001>$ direction in addition to 
a distortion in the
$<110>$ direction; alternatively, the suppression of the Co(1)-Cu(2)
peak requires a large Debye-Waller factor of $\sim 0.1$ \AA{}.
 This would produce the decreased amplitude
observed in the data, but we are unable to determine the exact nature of
the distortion at the present time.	

\acknowledgements

The experiments were
performed at the Stanford Synchrotron Radiation
Laboratory, which is operated by the U.S. Department of Energy, Division
of Chemical Sciences, and by the NIH, Biomedical Resource Technology 
Program, Division of Research Resources.  
The work is supported in part by NSF grant DMR-92-05204. 


\newpage

\newpage

\newpage

\newpage

\newpage

\newpage

\newpage

\clearpage
\onecolumn
\begin{table}[t]
\widetext

\caption{Single peak fit results of the  nearest neighbor oxygen atoms 
for both the
$c$-axis and $ab$-plane YBCO:Co and PBCO:Co thin film Co $K$-edge data.   
The uncertainties in
measurements were estimated to be  0.02 \AA{} for bond 
lengths, $R$,  and 0.005 \AA{} for the Debye-Waller
parameter, $\sigma$. The amplitude reduction factor, $S_0^2$, has
an average value of 0.7.  The Cu(1)/Co(1)--O
bond lengths should be selected appropriately: Cu(1) for YBCO and
Co(1) for YBCO:Co or PBCO:Co. }

\begin{tabular}{lddddd}
 & YBCO& \multicolumn{3}{c}{YBCO:Co} & PBCO:Co\\
\tableline
Concentration, x & 0 & 0.07 & 0.10 & 0.17 & 0.05 \\
\tableline
R (Cu(1)/Co(1)--O(4)) [\AA{}] & 1.8597\tablenotemark[1] 
	& 1.79 & 1.80 & 1.83 & 1.78 \\
& & & & & \\
R (Cu(1)/Co(1)--O(1)) [\AA{}] & 1.9407\tablenotemark[1]
	 & 1.86 & 1.88 & 1.88 & 1.87 \\
& & & & & \\
$\sigma$ (Co(1)--O(4)) [\AA{}] &  & 0.049& 0.058& 0.048& 0.040\\
& & & & & \\
$\sigma$ (Co(1)--O(1)) [\AA{}] &  & 0.041& 0.041& 0.069& 0.056\\
\end{tabular}
\label{table_result}
\tablenotetext[1] { Neutron diffraction results at $T$=80 K from
Sharma {\it et al.}\protect\cite{Sharma91}}

\end{table}

\end{document}